\documentclass[12pt, a4paper, final]{article}
\usepackage{amsmath}
\usepackage{amssymb}
\usepackage[english]{babel}
\usepackage{graphicx}
\usepackage{authblk}
\usepackage[colorlinks=true, linktocpage=false, draft=false]{hyperref}
\usepackage[doipre={doi:~}, citeurlpre={url:~}, citeurlpost={}]{uri}
\usepackage[left=3cm, right=1.5cm, top=2cm, bottom=2cm]{geometry}
\usepackage[style=gost-numeric-min, blockpunct=space, language=auto, autolang=other, sorting=none, bibdoi=true, biburl=true, movenames=false, maxnames=4, minnames=3, useeditor=true, backend=biber]{biblatex}

\AtEveryBibitem{\clearfield{month}}
\AtEveryBibitem{\clearfield{number}}
\AtEveryBibitem{\clearfield{issue}}
\DeclareFieldFormat[article]{title}{}
\DeclareFieldFormat{doi}{%
  doi\addcolon\space
  \ifhyperref
    {\href{https://doi.org/#1}{\nolinkurl{#1}}}
    {\nolinkurl{#1}}}
\DeclareFieldFormat{url}{%
  url\addcolon\space
  \ifhyperref
    {\href{#1}{\nolinkurl{#1}}}
    {\nolinkurl{#1}}}

\allowdisplaybreaks

\makeatletter
\predisplaypenalty=\@medpenalty
\makeatother

\title{Surface Ion-Sound Wave in Magnetic Arch With High Pressure Plasma}

\author[1,2]{S.~A.~Koryagin}
\author[1,2]{M.~E.~Viktorov}
\author[1,2]{A.~V.~Korzhimanov}
\author[1]{A.~A.~Elyasin}
\affil[1]{A.~V.~Gaponov-Grekhov Institute of Applied Physics of the Russian Academy of Sciences, Nizhny Novgorod, 603950 Russia}
\affil[2]{Lobachevsky University, Nizhny Novgorod, 603022 Russia}

\begin{document}
\linespread{2}

\maketitle

\abstract{The work analytically substantiates the parameters of the surface wave found in numerical modelling of the collision of two oncoming supersonic plasma flows inside a magnetic arc in application to the experiment on the laboratory setup ``Solar Wind'' (Inst. Appl. Phys RAS). An ion-acoustic surface wave exists in the regime of dense plasma flows when their dynamic pressure is of the order of the pressure of an undisturbed magnetic field, so that the flows push the initial magnetic field out of their volume. The wave frequency is in the range between the ion gyrofrequencies inside the plasma bundle and in the outer region of the confining magnetic field. In the external rarefied medium, the near-surface structure is a heterogeneous magnetic sound, consistent in pressure and low total polarisation of the medium with the ``isotropic'' ion sound confined from the inside in a dense plasma bundle. The energy of the structure is mainly contained in the kinetic energy of the wave motion of ions inside the tube. At the same time, the electric field strength is sharply increased outside. Firstly, the latter circumstance arises from the need to maintain a uniform electron electric drift velocity inside the transition layer. Secondly, the energetically weak ion sound propagating into the outer environment is close to electrostatic ion oscillations below the ion gyrofrequency in the external region, which are characterised by increased electric field strength across the ambient magnetic field.}

\clearpage

\subsection*{INTRODUCTION}

The compact experimental setup ``Solar Wind'' is designed to simulate plasma processes in a magnetic arch such as a coronal loop on the Sun: plasma confinement, system destruction with matter ejection when plasma pressure increases, and the generation of nonthermal radio emission \cite{Viktorov-2015-transl, Viktorov-2017}. The laboratory setup is distinguished by the creation of plasma not in a gas discharge, but by an arc discharge on a highly conducting metal cathode. The metal bases of the plasma loop make it difficult for the plasma to polarize across the magnetic field and, as a result, prevent possible deviation of the matter flow from the magnetic field line through electric drift in the polarization field of the medium \cite{Lindberg-1978}, and also contribute to the stabilization of the flute instability \cite[\S~14.10]{Mikhailovsky-II-book-transl}.

Arc discharges at the bases of the magnetic arch create plasma streams with thermal pressures lower than their dynamic pressures, corresponding to the supersonic motion of the jets. Varying the arc discharge current allows one to vary the plasma jet density and, consequently, its dynamic pressure. The latter can reach the pressure of an undisturbed vacuum magnetic field in the system. When the opposing streams collide at the apex of the arch, part of their energy from their translational motion is converted into thermal energy of cyclotron rotation. At the same time, the monovelocity streams broaden in longitudinal velocity, and their distribution can transform from a two-peaked, two-stream distribution to a distribution with a single maximum in longitudinal momentum. The experiment demonstrates both a stable plasma rope with low dynamic pressure from the jets and a scenario where the loop apex collapses, with a portion of the plasma stream being carried upward from the arch.

The experimental study is accompanied by numerical simulation of the collision of supersonic plasma flows in a magnetic arch~\cite{Korzhimanov-2025}. Two-dimensional simulation considers the arch as an infinitely wide curved layer, uniform along the direction $\boldsymbol{\tau}^0$ of its ``transverse'' translational symmetry (across the magnetic induction lines). This approach excludes flute instability as a possible mechanism of system destruction~\cite[\S~10.5]{Golant-book-eng}. Numerical simulation revealed that if the dynamic pressure of each flow is of the order of the magnetic pressure at the base of the arch, the plasma rope expels the magnetic field from its volume, simultaneously expanding in its cross section. As a result, after the initial transient process, the thermal pressure of the medium inside the arch layer significantly exceeds the local magnetic pressure in it. The stronger magnetic field expelled outward keeps the plasma from further expansion with its pressure.

Immediately after the flows meet at the apex of the arch, internal volumetric magnetohydrodynamic waves are excited in the system, in particular, of the ion-sonic type, which partially stop the relative motion of the flows and also redirect the hydrodynamic flow in the tube of variable cross-section. Such dynamics correspond to analytical estimates and numerical calculations \cite{Forslund-1970a, Forslund-1970b, McKee-1970}, according to which shock waves with a sharp difference in plasma density and directional velocity on either side of the front are formed in an unmagnetized system if the Mach number $M$ of each flow for the ion-sonic velocity $c_\text{s}$ exceeds unity, but remains below $3{.}0\text{--}3{.}5$. At higher Mach numbers, single-velocity ion flows only partially broaden in longitudinal velocity due to the generation of strong ion-acoustic waves such as kinks, so the ion velocity distribution retains a double-humped profile. In the experiment and calculations, the parameter $M=3.5$ corresponds to the theoretical boundary between the two deceleration regimes of opposing jets. 

In numerical simulations, at the steady-state stage, the upper and lower arch vaults are clearly distinguished by a spatially quasi-periodic electrical field, which is directed both along the system's homogeneity axis and normal to the arch layer. The structure of the electrical field resembles a surface standing wave, concentrated outside the arch vaults in a relatively narrow layer with a width of approximately the wavelength.

The purpose of the further presentation is to substantiate the existence of a surface wave of the ion-sound type on the vaults of the magnetic arch and to explain the increased electrical tension outside the tube.

\subsection*{1.~EXPERIMENTAL SETUP ``SOLAR WIND''}

In the experimental setup ``Solar Wind'' \cite{Viktorov-2015-transl, Viktorov-2017}, the magnetic arch is organized in a cylindrical vacuum chamber with a diameter of $183~\text{mm}$ and a width of $130~\text{mm}$. Two solenoids embrace two mutually perpendicular flange radial inputs to the chamber with a diameter of $65~\text{mm}$ and form an unperturbed arch similar to a quarter of a circle with a radius of 9~cm. The magnetic field at the center of the solenoids reaches 3{.}3~T at a maximum current in the coils of about $5{.}5~\text{kA}$, and at the center of the chamber and approximately at the top of the arch $0{.}12~\text{T}$ at the same current. The current pulse in the solenoids of 3~ms significantly exceeds the duration of the observed plasma processes, which allows us to consider the external magnetic induction to be stationary.

An arc discharge as a plasma source occurs on an aluminum cathode with a diameter of $10~\text{mm}$. The cathode is placed in front of each solenoid mirror-symmetrically to the entrance to the vacuum chamber relative to the center of the solenoid. This arrangement provides a plasma tube diameter at the entrance to the chamber similar to that at the cathode~--- $10~\text{mm}$. The arc discharge current $I_\text{gen}$ varies from $0{.}1$ to $7~\text{kA}$ and creates plasma with an ion density at the cathode $n_\text{i\,gen}=2\cdot10^{14}I_\text{gen}[\text{kA}]$ in the range $10^{13}\text{--}10^{15}~\text{cm}^{-3}$. The average charge number of ions $Z=1{.}7$. The plasma cloud front propagates from the cathode with a characteristic velocity of $v_\text{i}=15~\text{km}/\text{s}$, which is $M=3{.}5$~times greater than the ion-sound velocity $c_s=\sqrt{ZT_\text{e}/m_\text{i}}=4{.}3~\text{km}/\text{s}$ for the electron temperature $T_\text{e}=3~\text{eV}$ and the ion mass $m_\text{i}=27~\text{u}$. The arc discharge lasts $20~\mu\text{s}$ and creates a plasma bunch of length $30~\text{cm}$, which is $2{.}1$~times greater than the length of the arch loop.

The induction at the cathode and directly at the entrance to the vacuum chamber on the central magnetic field line of the arch is somewhat lower than at the center of the solenoids and exceeds the induction at the apex of the arch by approximately a mirror ratio of $\alpha=9~\text{times}$. At the entrance to the chamber, the undisturbed magnetic pressure greatly exceeds the dynamic pressure of the plasma. However, as we move from the base to the top of the arch, the magnetic pressure decreases faster than the plasma pressure, and for the characteristic induction at the top $B_\text{top}=0{.}1~\text{T}$ the pressure $p_\text{mag\,top}=B_\text{top}^2/(8\pi)=4{.}0\cdot10^4~\text{dyn}/\text{cm}^2$ turns out to be of the order of the total dynamic pressure of the two flows $p_\text{pl\,top}=2n_\text{i\,gen}m_\text{i}v_\text{i}^2/\alpha=2{.}3\cdot10^4~\text{dyn}/\text{cm}^2$ for the ion density in the discharge $n_\text{i\,gen}=10^{15}~\text{cm}^{-3}$.

\subsection*{2.~PARAMETERS OF NUMERICAL SIMULATION}

In the numerical simulations \cite{Korzhimanov-2025}, the hybrid calculation scheme AKA \cite{Sladkov-2020} was used, which solves the kinetic equation for ions and treats electrons hydrodynamically as a massless fluid. On the right-hand side of the electron Euler equation, the gradient of the thermal electron pressure is in balance with the ``total'' Lorentz force from the electric and magnetic fields. The electron pressure evolves along the hydrodynamic trajectory as in a local adiabatic process with anisotropic compression along and across the magnetic induction line (similar to the approximation in the Chew~--- Goldberger~--- Low theory). Maxwell's equations are considered in the Darwin approximation: the polarization current of each plasma fraction significantly exceeds the displacement current.

The spatial configuration in the two-dimensional numerical calculation qualitatively represents the experimental vacuum chamber and an arch, approximately three times larger in linear size in the sagittal cross-section, to reduce the effect of the non-zero Larmor radius of the ions. The three-dimensional arch is replaced by a two-dimensional structure resembling a wide arched layer, which qualitatively corresponds to a mirror-symmetric elongation of the three-dimensional tube into a layer along the normal to its sagittal cross-section. Due to the slower decay of the magnetic field of the external two-dimensional solenoids with distance, the magnetic induction at the beam injection point into the chamber is fixed in the range between the mirror and apex fields in the three-dimensional arch: $B_\text{foot}=0{.}25~\text{T}$. The initial flow velocity $v_\text{i}$ was varied between $10$ and $20~\text{km}/\text{s}$, which covers the experimental value of $15~\text{km}/\text{s}$. The plasma density at the injection point was increased approximately 10~fold to $n_\text{i\,foot}=10^{16}~\text{cm}^{-3}$, so that the dynamic pressure of one flow at the injection point $m_\text{i}n_\text{i\,foot}v_\text{i}^2$ was 1{.}8 and 7{.}2 magnetic pressure $B_\text{foot}^2/(8\pi)$ for the indicated values of hydrodynamic velocity.

In numerical simulations, a surface wave existed both in the case of single-stream injection (Fig. 1) and in the case of two jets (Fig. 2). Increasing flow velocity expanded the layer occupied by the surface wave (Fig. 3).

The presence of a surface wave even in the case of a single flow complicates the interpretation of the mechanism for generating this disturbance due to the two-flow instability for ion sound \cite[\S~3.5]{Mikhailovsky-I-book-transl}. At the same time, the plasma flow configuration under consideration is qualitatively similar to the solar wind flow around the Earth's magnetosphere, namely, the magnetopause region far from the head point of stagnation, where solar wind plasma, having passed the bow shock wave, in the form of a magnetosheath flows around the planetary magnetosphere, filled with a less dense medium \cite{Plaschke-2016-inbook, Archer-2024}. The excitation of surface magnetohydrodynamic waves at the magnetopause via the Kelvin-Helmholtz instability is considered as a possible mechanism for low-frequency Pc5 oscillations in the Earth's magnetosphere \cite{Plaschke-2011, Plaschke-2016-inbook, Pilipenko-2017-transl}. However, the Kelvin-Helmholtz instability criterion \cite[\S~106]{Chandrasekhar-book-1961}, \cite{Sen-1963} is not satisfied in numerical simulations due to the high Alfven velocity outside the plasma tube $c_\text{A\,ext}$ compared to the flow velocity~$v_\text{i}$.

\subsection*{3.~SURFACE ION-SOUND WAVE}

\subsection*{3.1.~Conditions for the existence of a surface ion-acoustic wave}

Two-dimensional modeling excludes from its consideration surface and drift magnetohydrodynamic waves, which require for their existence a non-zero component $k_\tau$ of the wave vector along the homogeneity axis~$\boldsymbol{\tau}^0$, in particular, the so-called Kruskal~--- Schwarzschild waves \cite{Kruscal-1954, Plaschke-2016-inbook}, otherwise known as surface Alfven waves \cite{Goossens-2012} due to the relatively weak compressibility of the plasma in them. As a consequence, the flute instability to which the indicated waves are subject \cite[\S~10.5]{Golant-book-eng}, \cite{Kruscal-1954}, and the buildup of the near-surface ion-sound wave by the diamagnetic current of thermal electrons inside the magnetic wall of the medium \cite[\S~9.2]{AlexandrovBogdankevichRukhadze-book-eng} are excluded.

Wentzel \cite{Wentzel-1979a} showed that a surface magnetohydrodynamic wave with a zero component $k_\tau$ of the wave vector is absent in a low-pressure $p_\text{pl}\ll B^2/(8\pi)$ plasma placed in a quasi-uniform magnetic field~$\mathbf{B}$. The thickness of the wave structure increases with decreasing component $k_\tau\to0$, so that the surface wave becomes indistinguishable from the bulk wave. However, a wave with component $k_\tau=0$ pressed to the boundary is preserved in the case of a high-pressure plasma that has forced out the magnetic field that confines it. The latter case corresponds to the conditions of the numerical simulation discussed.

The magnetohydrodynamic approach justifiably neglects the displacement current, in particular in the discussion of \cite{Wentzel-1979a}. However, taking into account the displacement current outside the plasma half-space, for example in a vacuum, preserves the surface magnetohydrodynamic wave of the ion-acoustic type even at a low ratio of the plasma $p_\text{pl}$ and magnetic $p_\text{mag}=B^2/(8\pi)$ pressures in the frequency range $\omega_{B\text{i}}<\omega\ll\mathop{\mathrm{min}}(\omega_{B\text{e}},\omega_\text{pi})$~--- below the electron gyrofrequency $\omega_{B\text{e}}$ and the ion plasma frequency~$\omega_\text{pi}$,~--- when the electrons are magnetized and the transverse permittivity of the ion fraction is negative \cite[\S~2.4]{Kondratenko-book-1985-eng}. The thickness of the near-wall wave structure in the plasma $1/\varkappa_\text{pl}$ exceeds the wavelength along the boundary $2\pi/k_{\parallel}$ by a value of the order of the large absolute value of the transverse dielectric susceptibility of the ion fraction $\omega_\text{pi}^2/(4\pi\omega^2)\gg1$~\cite{Azarenkov-1998-transl}.

\subsection*{3.2.~Choosing an electrodynamic model of the external environment}

In turn, under the conditions of the ``Solar Wind'' setup, the high permittivity $|\omega_\text{pi}^2/(4\pi\omega^2)|_{\omega\sim\omega_{B\text{i}}}\sim4\cdot10^8\gg1$ excludes the vacuum approximation outside the plasma tube, since it is difficult to expect a plasma density difference of $10^8$~times inside the magnetic wall at the upper and lower vaults of the arch for the plasma frequency outside the tube to drop below the local ion gyrofrequency. Therefore, in the electrodynamic problem of a surface wave, the approximation of a rarefied plasma, rather than a vacuum outside the arch layer, seems more adequate. In this case, the pressed vacuum electrostatic field from the wave surface charge at the plasma boundary should be replaced by the field of an electrostatic ion-acoustic wave outside the arch layer (pressed or outgoing volume). A volumetric electrostatic wave carries away energy from the near-surface structure, leading to its slow attenuation. The escape energy flux decreases proportionally to the low density of the external medium.

\subsection*{3.3.~Set of boundary conditions}

Plasma gyrotropy in a magnetic field violates the mirror symmetry of the problem relative to the plane of wave incidence on the boundary. As a result, the electrodynamic boundary conditions at the interface of the media include the ``continuity'' of four tangential field components: two components each for the electric intensity and the magnetic induction. In the case of high-pressure plasma ($p_\text{pl}\sim p_\text{mag}$), the boundary between the media itself acquires a corrugation corresponding to the wave motion.

The ``continuity'' of the longitudinal tangential wave component of magnetic induction is transformed into a condition for the variable jump in magnetic field magnitude upon crossing the boundary, which takes into account the oscillating surface diamagnetic current due to the difference in plasma pressures in the contacting half-spaces. Given the relationship between the diamagnetic current and the plasma pressure gradient, the condition under discussion takes the form of a balance between the total plasma and magnetic pressures on either side of the boundary.

In the processes under consideration, with a frequency below the electron gyrofrequency, the transverse motion of electrons represents an electrical drift in an alternating field, which corresponds to the mutual freezing of the electron fraction and magnetic induction. Therefore, by the interface of the media, we mean the boundary between the electron fractions and the magnetic surface frozen into it. At the moving boundary, the tangential component of the electric intensity (simultaneously transverse to the unperturbed magnetic induction) undergoes a jump (in the laboratory frame of reference), which reflects the non-zero circulation of the electric field in the case of different magnitudes of the unperturbed magnetic field on opposite sides of the moving contact between the media. The discussed difference in the ``tangential-transverse'' electrical field strength becomes zero in the moving frame of reference, instantaneously following the boundary along the normal at the velocity of the electron electrical drift (as a result of the standard conversion of electromagnetic field vectors between the laboratory and moving frames). Therefore, the boundary condition of ``continuity'' of the ``tangential-transversal'' electrical intensity on the moving boundary is reduced to the condition of equal displacement of electrons in the electrical drift on both sides of the boundary of the media.

In turn, the continuity of the tangential wave component of the magnetic induction, which is simultaneously orthogonal to the unperturbed induction $\mathbf{B}$, is equivalent to the continuity of the normal component of the electric induction $D_n=(\mathrm{i}c/\omega)\,(\mathbf{n}^0,\mathop{\mathrm{rot}}\mathbf{B})$ (in the particular case under consideration of the wave vector concentrated strictly in the plane of the vector $\mathbf{B}$, and the normal~$\mathbf{n}^0$ to the boundary~--- $k_\tau=0$). Non-zero electric induction (polarization) occurs from the relative displacement of plasma fractions, when the motion of ions is not reduced to an electric drift together with electrons.

Thus, in the following presentation, four conditions will be used at the interface of the media in the form of: 1) balance of magnetic and plasma pressures; 2) equal movement of electrons in electrical drift from opposite sides of the interface; 3) continuity of the normal component of electrical induction (polarization vector); 4) continuity of the longitudinal component of electrical intensity (along the magnetic field line).

The magnetohydrodynamic consideration \cite{Wentzel-1979a} did not take into account electrodynamic conditions 3 and 4 and thus missed the outgoing ion-acoustic wave, and consequently, this mechanism of surface structure attenuation.\,\footnote{\ At the same time, in the work \cite{Wentzel-1979b} a variant of surface wave dissipation due to its resonance with the Alfven wave inside a magnetic wall of non-zero thickness~$a$ was proposed. The decrement is of the order of $\omega\,(k_{\parallel}a)$. We neglect this resonance.} In turn, the electrodynamic approach \cite{Kondratenko-book-1985-eng} is valid for a medium with low plasma pressure ($p_\text{pl}\ll p_\text{mag}$), when electrons are unable to deform the field frozen in them, corrugate the intermedium boundary and drift behind ions, nullifying the transverse polarization of the medium. The limit of solid magnetic field lines allows localization of the near-surface structure in the plasma only on a wide spatial scale compared to the longitudinal wavelength at low frequencies $\omega\ll\omega_\text{pi}$. This version does not include the structure discovered in the numerical simulation of \cite{Korzhimanov-2025}, where the pressures $p_\text{pl}$ and $p_\text{mag}$ are of the same order of magnitude.

At the same time, four boundary conditions determine the number of waves involved in the subsequent analysis~--- four. The strong difference in plasma densities on either side of the boundary allows us to resolve the system of boundary conditions on the wave amplitudes using successive approximations: the linear system of equations is essentially solved by the Gaussian method, and the standard condition of the system's zero determinant reduces to the solvability of the pressure balance equation, where all quantities are expressed through a single variable~--- the potential of the pressed ion-acoustic mode in a dense plasma.

\subsection*{4.~MODEL OF A PLASMA LAYER AS A HALF-SPACE THAT FORCED OUT THE MAGNETIC FIELD}

Based on the described qualitative scenario, we formulate a specific model of the boundary between media for calculating the surface wave parameters. Two isothermal plasma half-spaces with very different unperturbed ion densities $n_\text{in}$ and $n_\text{ext}\ll n_\text{in}$ touch along a plane on which the stationary part of the magnetic field induction transitions from a ``low'' value $B_\text{in}$ inside the denser medium to a high value $B_\text{ext}\gg B_\text{in}$ outside. The electron pressure inside a dense medium $p_\text{e\,in}=Zn_\text{in}T_\text{e}$ significantly exceeds the local magnetic pressure $p_\text{mag\,in}=B_\text{in}^2/(8\pi)$, where $Z$~is the charge number of ions, $T_\text{e}$~is the electron temperature. We assume that ions are colder than electrons for ion sound to exist. The dense internal plasma is kept from expanding by the magnetic pressure in the rarefied external medium: $B_\text{ext}^2/(8\pi)\approx p_\text{e\,in}$. For the given ratio of pressures and plasma densities, the uniform ion-sound velocity $c_\text{s}=\sqrt{ZT_\text{e}/m_\text{i}}$ significantly exceeds the Alfven velocity in the inner half-space $c_\text{A\,in}=B_\text{in}/\sqrt{4\pi m_\text{i}n_\text{in}}$ and, on the contrary, is significantly lower than the similar velocity in the outer region $c_\text{A\,ext}=B_\text{ext}/\sqrt{4\pi m_\text{i}n_\text{in}}$\,, where $m_\text{i}$~--- the ion mass:
\begin{equation}
c_\text{A\,in}\ll c_\text{s}\ll c_\text{A\,ext}.
\label{SequenceOfIonSoundAndAlfvenVelocities}
\end{equation}

The problem being solved generally relates to the Fresnel problem of the reflection and transmission coefficients of a wave colliding with a flat interface (with a thickness less than the transverse lengths of the incident, reflected, and transmitted waves). The sought-after surface structure in a dense plasma corresponds to the regime of total internal reflection for ion sound incident on the interface from a rarefied half-space. The incident wave pumps the surface structure to a steady-state level over a large number of field periods (characterizing the Q factor of the pressed mode). Upon reaching the steady-state regime, the energy of the incident ion sound is completely converted into the reflected wave. We interpret the problem not as pumping and steady-state in the Fresnel problem, but, instead, as the slow decay of the surface structure after the incident ion sound is switched off (in the absence of internal pumping from the nonequilibrium two-stream ion fraction).

A surface wave should be expected at frequencies $\omega$ exceeding the ion gyrofrequency in a half-space with a weak magnetic field $\omega_{B\text{i\,in}}$. At lower frequencies ($\omega<\omega_{B\text{i\,in}}$), the electrons cease to compensate the ion current along each Cartesian coordinate and provide only zero divergence of the total electric current, somewhat similar to the unipolar diffusion regime. Under such conditions, there is a non-zero wave polarization of the dense medium across the magnetic field and it is not possible to ensure the continuity of the normal component of the electric induction at the boundary between the half-spaces, since the diagonal elements of the transverse ionic dielectric susceptibility acquire the same sign on both sides of the boundary, while the off-diagonal elements of the dielectric constant tensor vanish due to the joint electric drift of all plasma particles in the magnetic field. 

In turn, the existence of a weakly damped surface structure in the ``high-frequency'' limit of $\omega\gg\omega_{B\text{i\,ext}}$ also faces a significant limitation in the form of the same (already negative) sign of the transverse ionic polarizability in both half-spaces. In ionic sound in a rarefied half-space, the separate motion of ions and electrons is still preserved due to the low electron pressure compared to the magnetic pressure ($c_\text{s}^2\ll c_{\text{A}\,ext}^2$), unlike in a dense half-space, where the opposite relationship between pressures is fulfilled and electrons neutralize the ion current along each Cartesian coordinate (see the next subsection 4.1). The latter creates zero electric induction in ionic sound, and not only its zero divergence. 

The joint motion of electrons with the frozen boundary in both half-spaces (as well as the balance of pressure perturbations at the boundary) can be ensured by a whistler pressed to the boundary in the dense half-space (the continuation of fast magnetic sound into the frequency range $\omega>\omega_{B\text{i\,ext}}$). However, the continuity of the normal component of zero electric induction at the boundary requires the joint motion with the interface of the media of not only the electron but also the ion fraction in the rarefied half-space. The required dynamics is achieved only if the ion sound in the rarefied half-space does not decrease, but increases with distance from the boundary~--- the wave structure is not localized near the interface of the media in the range $\omega\gg\omega_{B\text{i}\,ext}$.

Thus, we restrict ourselves to the range between the ion gyrofrequencies in the dense and rarefied half-spaces:
\begin{equation}
\omega_{B\text{i\,in}}\ll\omega\ll\omega_{B\text{i\,ext}}.
\label{FrequencyBandBetweenIonGyrofrequencies}
\end{equation}
This interval is distinguished by the existence of modes with the joint motion of both plasma fractions in each half-space (and, consequently, a small electric field induction compared to the polarization of each fraction): ion sound in the dense half-space and magnetic sound in the rarefied region. The phase velocity of each wave (in the regime of purely real components of the wave vector) is of the order of the maximum of the ion-sound and Alfven velocities for its region. Therefore, both modes are classified as fast magnetosonic waves in their half-spaces.

\subsection*{4.1.~Mutual compensation of ion and electron currents in ion sound in a half-space with dense plasma}

An ion-acoustic type electrostatic wave with potential $\Phi_\text{in\,IS}(\mathbf{r})$ establishes a Boltzmann distribution of electrons, which corresponds to a perturbation of the plasma pressure
\begin{equation}
\delta p_\text{in}=n_\text{in}e\Phi_\text{in\,IS}.
\label{PlasmaPressureDisturbanceInDenseHalfSpace}
\end{equation}
In this case, the electron current from the electric drift in the transverse electric field $E_n=-\partial\Phi_\text{in\,IS}/\partial x$ is compensated by the diamagnetic current from the electron pressure gradient $\partial p_\text{e\,in}/\partial x$. From here on, the $x$ axis is directed along the normal $\mathbf{n}^0$ to the flat boundary of the media from the dense to the rarefied half-space. The transverse homogeneity axis of the problem $\boldsymbol{\tau}^0=[\mathbf{b}^0,\mathbf{n}^0]$ forms a right triple with the direction $\mathbf{b}^0$ of the stationary magnetic induction $\mathbf{B}$ and the unit vector $\mathbf{n}^0$.

Ionic sound in a magnetic field is not a purely electrostatic wave due to the anisotropy of the medium, at least of the electron fraction, and contains a weaker inductive electric field. In the frequency range (\ref {FrequencyBandBetweenIonGyrofrequencies}) and at higher frequencies up to $\omega\sim\omega_{B\text{i\,ext}}^2/(2\omega_{B\text{i\,in}})$, the off-diagonal elements of the permittivity tensor in the dense half-space $\varepsilon_{n\tau}=-\varepsilon_{\tau n}\approx\mathrm{i}\omega_\text{pi\,in}^2/(\omega\omega_{B\text{i\,in}})=\mathrm{i}\,[c^2/(2c_\text{s}^2)]\,[\omega_{B\text{i\,ext}}^2/(\omega\omega_{B\text{i\,in}})]$, describing the current of electric drift of electrons for the process with time dependence $\exp(-\mathrm{i}\omega t)$, exceed in modulus not only the diagonal elements $\varepsilon_{nn}=\varepsilon_{\tau\tau}\approx-\omega_\text{pi\,in}^2/\omega^2$ of the same tensor from the ion polarization current, but and the square of the conditional refractive index for ionic sound $c^2/c_\text{s}^2$. Under the conditions of the specified dominance of off-diagonal elements, the component of the ion polarization current transverse to the magnetic field $j_{\text{i}\,n}=\mathrm{i}\omega_\text{pi\,in}^2/(4\pi\omega)\,E_n$ is compensated by the electron current of the electric drift
\begin{equation}
j_{\text{e}\,n}=\omega_\text{pi\,in}^2/(4\pi\omega_{B\text{i\,in}})\,E_\tau
\label{NormalElectronCurrentFromElectricDrift}
\end{equation}
in the inductive electric field~$E_\tau=-\mathrm{i}\omega_{B\text{i\,in}}E_n/\omega\ll E_n$.\,\footnote{\ In a rarefied half-space at frequencies $\omega>\omega_{B\text{i\,ext}}$, the off-diagonal elements of the dielectric tensor are below the square of refractive index for ionic sound, which ultimately does not allow generating a sufficient induction field $E_\tau$ and compensating for the ionic polarization along each Cartesian coordinate by the electron electric drift current.} The latter is directed along the translational symmetry axis of the problem $\boldsymbol{\tau}^0=[\mathbf{b}^0,\mathbf{n}^0]$. In this regime, the magnetic field lines are frozen into the electron, and consequently, the ionic fractions. 

The electric field strength $E_\tau$ along the translational symmetry axis in the absence of other Cartesian components of the electric field in the frequency range (\ref{FrequencyBandBetweenIonGyrofrequencies}) itself creates an electron current with zero divergence, so as not to disturb the electron density, and therefore the electron pressure, and thus not to create an electric field strength along the undisturbed magnetic field. Therefore, the drift current component (\ref{NormalElectronCurrentFromElectricDrift}) is accompanied by a longitudinal electron current
\begin{equation}
\tilde{j}_{\text{e}\,\parallel}=-(k_n/k_{\parallel})\,j_{\text{e}\,n}=\mathrm{i}\omega_\text{pi\,in}^2k_n/(4\pi\omega k_{\parallel})\,E_n.
\label{LongitudinalElectronCurrentNotToDisturbElectronPressure}
\end{equation}
It corresponds to nonzero elements of the dielectric tensor $\varepsilon_{\tau\parallel}=-\varepsilon_{\parallel\tau}=(k_n/k_{\parallel})\,\varepsilon_{n\tau}$ for processes with a phase velocity below the thermal velocity of electrons~\cite[f.~(4.98)]{Zheleznyakov-book-eng}. Here $k_n$ and $k_\parallel$ are components of the wave vector $\mathbf{k}=k_n\mathbf{n}^0+k_{\parallel}\mathbf{b}^0$.

The condition of zero total longitudinal electron current (\ref{LongitudinalElectronCurrentNotToDisturbElectronPressure}), polarization ion current $j_{\text{i}\,\parallel}=\mathrm{i}\omega_\text{pi\,in}^2/(4\pi\omega)\,E_{\parallel}$ and electron current of longitudinal Debye screening $j_{\text{e}\,\parallel}=-\mathrm{i}\omega\omega_\text{pi\,in}^2/(c_\text{s}^2k_{\parallel}^2)\,E_{\parallel}$ taking into account the potentiality of the components $E_n=k_nE_{\parallel}/k_{\parallel}$ of the electric field determines the dispersion relation for ion sound, as in unmagnetized plasma:
\begin{equation}
\omega^2=c_\text{s}^2\,(k_{\parallel}^2+k_n^2).
\label{DispersionRelationForIonSound}
\end{equation}

For an ion sound pressed against a boundary in a dense half-space, the wave vector component
\begin{equation}
k_n=-\mathrm{i}\varkappa_\text{in\,IS}
\label{ImaginaryNormalComponentOfIonSoundWaveVector}
\end{equation}
is purely imaginary (neglecting the temporal attenuation of the wave structure). Therefore, the frequency of the surface wave is lower than the frequency of the bulk sound for the same longitudinal wavenumber~$k_{\parallel}$.

\subsection*{4.2.~Electric field in the transition layer between plasma half-spaces}

In the transition layer between the plasma half-spaces, the transverse diagonal elements of the permittivity tensor $\varepsilon_{nn}=\varepsilon_{\tau\tau}=\omega_\text{pi}^2/(\omega_{B\text{i}}^2-\omega^2)$ and the off-diagonal elements $\varepsilon_{n\tau}=-\varepsilon_{\tau n}=-\mathrm{i}\,\omega_\text{pi}^2\omega/[\omega_{B\text{i}}\,(\omega_{B\text{i}}^2-\omega^2)]$, which describes the electric drift current of ions and electrons, exhibit a resonant feature on the plane where the local ion gyrofrequency $\omega_{B\text{i}}$ coincides with sound frequency~$\omega$. To maintain a continuous zero normal component of the electric induction in the transition layer, as in the dense half-space, the normal $E_n$ and tangential $E_\tau$ electric field strengths are related by the ``impedance'' equality
\begin{equation}
E_\tau=-\mathrm{i}\omega_{B\text{i}}E_n/\omega.
\label{ImpedanceRelationForNormalAndTangentialTensionInTransitionalLayer}
\end{equation}
At the same time, to maintain a uniform electron drift velocity $cE_\tau/B$ within the transition layer, the $E_\tau$ component must vary linearly proportional to the magnetic induction $B\propto\omega_{B\text{i}}$. Therefore, the induction component $E_\tau$ increases by $\omega_{B\text{i\,ext}}/\omega_{B\text{i\,in}}\gg1$ times when passing through the boundary into the rarefied half-space and becomes significantly greater than the component~$E_n$. Thus, the homogeneity of the electron drift velocity in the boundary layer ensures the transformation of ion sound in the dense half-space into inhomogeneous magnetic sound in the rarefied half-space (TE wave), which is predominantly polarized along the translational symmetry axis of the problem in the range (\ref{FrequencyBandBetweenIonGyrofrequencies}).

The jump of the tangential component $E_\tau$ at the boundary between the half-spaces corresponds to the Faraday equation of electromagnetic induction for a moving boundary, if the difference in the stationary magnetic field and the displacement of the boundary with the electric drift velocity $cE_\tau/B$ are taken into account at the latter. At the same time, according to the impedance equality (\ref{ImpedanceRelationForNormalAndTangentialTensionInTransitionalLayer}), directly on the resonant plane $\omega_{B\text{i}}=\omega$, the polarization of the transverse electric field $E_n\mathbf{n}^0+E_\tau\boldsymbol{\tau}^0$ is circular in the direction of electron gyrorotation, which excludes resonance of the field with ions.

\subsection*{4.3.~Continuity of the normal component of electric induction at the boundary. Outgoing ion sound and whistler}

If we consider the surface wave problem as a Fresnel problem, where ionic sound is incident on a boundary from a rarefied half-space and reflected under conditions of total internal reflection, then the continuity condition at the boundary of the longitudinal component of the electric field strength and the zero normal component of the electric induction determine reflection with a unit coefficient in the amplitude of the electric potential. At the boundary, the electric potential in the incident and reflected waves is half the potential of the pressed ionic sound~$\Phi_\text{in\,IS}$ in the dense half-space. In turn, the nonzero normal components of the electric induction in the incident and reflected waves mutually compensate each other at the boundary of the half-spaces.

In the absence of an incident wave, the reflected wave alone no longer ensures the simultaneous continuity of both of the above-mentioned field components. Therefore, the outgoing ion-sound wave must be complemented by an outgoing whistler in the dense half-space in the considered version of the problem. The sharp difference in plasma density in the half-spaces requires a larger particle oscillation amplitude in the outgoing ion-sound wave than in the whistler wave to ensure equal normal components of the electric induction in them on either side of the boundary. Therefore, the amplitude of the electric field strength in the outgoing ion-sound wave is significantly higher than the whistler amplitude, allowing us to find the runaway wave amplitudes through successive iterations.

\subsection*{4.4.~Inhomogeneous magnetic sound in a rarefied half-space}

In the initial approximation, we neglect the outgoing ionic sound and whistler in order to determine the purely real frequency $\omega$ of the surface structure in the absence of radiation losses of waves. The dispersion relation for magnetic sound in a rarefied half-space
\begin{equation}
\omega^2=c_\text{A\,ext}^2\,(k_{\parallel}^2+k_n^2)
\label{DispersionRelationForMagneticSound}
\end{equation}
sets the purely imaginary normal component of the wave vector $k_n\approx\mathrm{i}\,|k_{\parallel}|$ to be approximately equal in absolute value to the longitudinal component $k_{\parallel}$ for the pressed TE mode in a rarefied half-space, since the Alfven velocity $c_\text{A\,ext}$ significantly exceeds the sound velocity $c_\text{s}$, and the single frequency (\ref{DispersionRelationForIonSound}) of the wave structure is of the order of or lower than $c_\text{s}k_{\parallel}\ll c_\text{A\,ext}k_{\parallel}$.

The condition of freezing the magnetic field into the electron fraction connects the perturbation of the longitudinal component of magnetic induction with the inhomogeneity of the normal component of the electron displacement $\xi_{\text{e}\,n}$ by the standard relation
\[
\delta B_{\parallel}=-\mathrm{i}k_n\xi_{\text{e}\,n}B_\text{ext}=|k_{\parallel}|\,\xi_{\text{e}\,n}
\]
\cite[f. (2.12)]{Kadomtsev-book-eng}. In turn, the balance of perturbations of magnetic
\begin{equation}
\delta p_\text{mag}=B_\text{ext}\,\delta B_{\parallel}/(4\pi)=2p_\text{mag\,ext}\,|k_{\parallel}|\,\xi_{\text{e}\,n}
\label{MagneticPressureDisturbance}
\end{equation}
and plasma (\ref{PlasmaPressureDisturbanceInDenseHalfSpace}) pressures on opposite sides of the boundary imposes an additional relationship between the wave numbers $k_{\parallel}$, $\varkappa_\text{in\,IS}$, and the frequency $\omega$ for the surface structure
\begin{equation}
2\,|k_{\parallel}|\,\varkappa_\text{in\,IS}\,\frac{c_\text{s}^2}{\omega^2}=1,
\label{PressureBalanceAsAdditionalSpectralRelation}
\end{equation}
if the displacement of electrons $\xi_{\text{e}\,n}$ at the boundary is taken to be only their displacement in the pressed ion-acoustic wave $\xi_{\text{e}n0}=Ze\,\varkappa_\text{in\,IS}\Phi_\text{in\,IS}/(m_\text{i}\omega^2)$.

The pressure balance in the form (\ref{PressureBalanceAsAdditionalSpectralRelation}) is expanded into explicit dispersion relations
\begin{gather}
\label{SpatialDecrementOfIonSoundWaveInDenseHemiSpace}
\varkappa_\text{in\,IS}=(\sqrt{2}-1)\,|k_{\parallel}|;\\
\label{SquaredFrequencyOfIonSoundInDenseHemiSpace}
\omega^2=2\,(\sqrt{2}-1)\,c_\text{s}^2k_{\parallel}^2\approx0.91^2c_\text{s}^2k_{\parallel}^2
\end{gather}
taking into account the general relation (\ref{DispersionRelationForIonSound}) between parameters $k_{\parallel}$, $\varkappa_\text{in\,IS}$ and $\omega$.

The same movement of electrons together with the boundary frozen in them both in the pressed ion sound $\xi_{\text{e}n0}=Ze\,\varkappa_\text{in\,IS}\Phi_\text{in\,IS}/(m_\text{i}\omega^2)$ in a dense half-space and in the inhomogeneous magnetic sound $\xi_{\text{e}n0}=\mathrm{i}ZeE_{\tau\,\text{MS}}/(m_\text{i}\omega_{B\text{i\,ext}}\omega)$ in a rarefied plasma explains the sharp increase in electrical intensity
\begin{equation}
E_{\tau\,\text{MS}}=\frac{\mathrm{i}\omega_{B\text{i\,ext}}}{\omega}\,
(-\varkappa_\text{in\,IS}\Phi_\text{in\,IS})
=(\sqrt{2}-1)\,\frac{\omega_{B\text{i\,ext}}}{\omega}\,(-\mathrm{i}\,|k_{\parallel}|\,\Phi_\text{in\,IS})
\label{TangentialElectricTensionOnOuterSurfaceOfPlasmaBoundary}
\end{equation}
outside the magnetic arc compared to the field inside~$\mathbf{E}_\text{in\,IS}=-(\varkappa_\text{in\,IS}\mathbf{n}^0+\mathrm{i}k_{\parallel}\mathbf{b}^0)\,\Phi_\text{in\,IS}$~--- by the order of $\omega_{B\text{i\,ext}}/\omega\gg1$ times. This circumstance corresponds to the growth of the $E_\tau$ component inside the transition layer, discussed in section~4.2. The increased $E_\tau$ strength outside the magnetic arch exactly corresponds to the result of the discussed numerical simulation~\cite{Korzhimanov-2025}.

\subsection*{4.5.~Outgoing ion sound}

The outgoing ion-sound wave in a rarefied half-space arises from the continuity of the longitudinal electrical field at the boundary of the half-spaces, neglecting the weaker field of the outgoing whistler. This circumstance imposes identical electrical potentials at the boundary in the pressed ion-sound structure in the dense half-space and the outgoing ion-sound wave.
 
In the frequency range (\ref{FrequencyBandBetweenIonGyrofrequencies}), the dispersion relation of ion sound in a rarefied half-space
\begin{equation}
\omega^2=\frac{c_\text{s}^2k_{\parallel}^2}{1+c_\text{s}^2k_n^2/\omega_{B\text{i\,ext}}^2}
\label{DispersionRelationForIonSoundInDenseHemiSpace}
\end{equation}
fixes the frequency-independent normal component of the wave vector of the outgoing ion sound
\begin{equation}
k_{n\,\text{ext\,IS}}^2=\frac{\sqrt{2}-1}{2}\,\frac{\omega_{B\text{i\,ext}}^2}{c_\text{s}^2}\gg k_{\parallel}^2,
\label{SquaredNormalWaveNumberForOutgoingIonSound}
\end{equation}
so that the frequency (\ref{DispersionRelationForIonSoundInDenseHemiSpace}) of the latter coincides with the frequency (\ref{SquaredFrequencyOfIonSoundInDenseHemiSpace}) of the pressed wave structure on the opposite side of the boundary. The normal component $v_{\text{gr}\,n}=\partial\omega/\partial k_n$ of the group velocity of the outgoing ion sound must be directed outward from the boundary of the half-spaces. The dispersion relation (\ref{DispersionRelationForIonSoundInDenseHemiSpace}) establishes the opposite directions of the normal components of the wave vector~$k_n$ and the group velocity~$v_{\text{gr}\,n}$. Therefore, the component $k_{n\,\text{ext\,IS}}$ of the wave vector of the outgoing sound is oriented not outward, but, on the contrary, inward of the dense half-space: $k_{n\,\text{ext\,IS}}=-|k_{n\,\text{ext\,IS}}|\,$.

The wavelength of the outgoing ion sound $2\pi/|k_{n\,\text{ext\,IS}}|$ is significantly shorter than the spatial scale of localization of the inhomogeneous magnetic sound near the boundary $1/|k_{\parallel}|$ (see formula (\ref{SquaredNormalWaveNumberForOutgoingIonSound})), which means smaller-scale variations in the normal component of the electric field $E_n$ than in the tangential component $E_\tau$. This ratio of scales qualitatively corresponds to the result of numerical simulation~\cite{Korzhimanov-2025}.

A large value of the wave number $|k_{n\,\text{ext\,IS}}|$ compared to the components $|k_{\parallel}|$ and $\varkappa_\text{in\,IS}$ for the pressed ion sound in a dense half-space (see expressions~(\ref{SpatialDecrementOfIonSoundWaveInDenseHemiSpace}) and (\ref{SquaredNormalWaveNumberForOutgoingIonSound})) leads to an increase in the normal component of the electric field directly outside the arch
\[
E_{n\,\text{ext\,IS}}\equiv\mathrm{i}\,|k_{n\,\text{ext\,IS}}|\,\Phi_\text{in\,IS}
=-(\sqrt{2}-1)\,\frac{\omega_{B\text{i\,ext}}}{\omega}\,(-\mathrm{i}\,|k_{\parallel}|\,\Phi_\text{in\,IS})
=-E_{\tau\,\text{MS}}
\]
up to the level of the tangential component (\ref{TangentialElectricTensionOnOuterSurfaceOfPlasmaBoundary}). The comparable level of normal and tangential components of electrical field strength outside the plasma column also corresponds to the result of numerical simulation~\cite{Korzhimanov-2025}.

\subsection*{4.6.~Departing whistler}

Unlike the pressed sound structure in a dense half-space, in the outgoing ionic sound there is no possibility of neutralizing the normal component of the alternating ion current by means of the electron current of the electric drift, since the latter equally involves both plasma fractions in the frequency range (\ref {FrequencyBandBetweenIonGyrofrequencies}). As a result, the uncompensated movement of the ion fraction from the polarization drift~\cite[Appendix I, \S~8.2]{Krall-book-eng} in the outgoing ionic sound induces an alternating surface charge at the boundary of the half-spaces
\begin{equation}
\sigma_\text{ext\,IS}=-Zen_\text{i\,ext}
\frac{\mathrm{i}\,|k_{n\,\text{ext\,IS}}|}{\varkappa_\text{in\,IS}}\,
\frac{\omega^2}{\omega_{B\text{i\,ext}}^2}\,
\xi_{\text{e}n0}.
\label{SurfaceChargeFromOutgoingIonSound}
\end{equation}

The charge (\ref{SurfaceChargeFromOutgoingIonSound}) at the interface is neutralized by the nonzero displacement of electrons relative to ions in the dense half-space due to motion in the field of the ``short-wave'' outgoing whistler, which restores the continuity of the normal component of the electric induction at the boundary. The dispersion equation for the whistler
\[
\omega=c^2k_{\parallel}k\omega_{B\text{i\,in}}/\omega_\text{pi\,in}^2
\]
aligns its wave vector almost perpendicular to the boundary between the half-spaces with a value
\[
k_{\text{W}\perp}\approx k_\text{W}=\sqrt{2\,(\sqrt{2}-1)}\,
\frac{\omega_\text{pi\,in}^2c_\text{s}}{c^2\omega_{B\text{i\,in}}}
=\sqrt{(\sqrt{2}-1)/2}\,
\frac{\omega_{B\text{i\,ext}}}{c_\text{s}}\,
\frac{\omega_{B\text{i\,ext}}}{\omega_{B\text{i\,in}}}\,.
\]
The whistler wavelength $2\pi/k_{\text{W}\perp}$ turns out to be significantly shorter than not only the characteristic spatial scale of the pressed ionic and magnetic sound $1/|k_{\parallel}|$, but also the wavelength of the outgoing sound in the rarefied half-space~--- by $\omega_{B\text{i\,ext}}/\omega_{B\text{i\,in}}\gg 1$ times (see formula (\ref{SquaredNormalWaveNumberForOutgoingIonSound})).

In contrast to low-pressure plasma ($c_\text{s}^2\ll c_\text{A}^2$), the whistler under discussion is polarized practically in the plane of the boundary between the half-spaces; the polarization is close to circular and contains approximately equal electric components along the magnetic field and along the translational symmetry axis $\boldsymbol{\tau}^0$, while the normal component of the electric field is significantly lower. This polarization feature occurs primarily because the whistler phase velocity is lower than the thermal electron velocity, so that the electric field strength along the translational symmetry axis excites an electric drift current together with a longitudinal current without perturbing the electron density (we are talking about the current (\ref{LongitudinalElectronCurrentNotToDisturbElectronPressure}) in the presence of only the field strength $E_\tau$ in the absence of the component $E_n$). At the same time, the diagonal elements of the plasma dielectric susceptibility, including those along the external magnetic field, are of the same order and significantly lower than the square of the whistler refractive index $k_{\perp}^2c^2/\omega^2$.

Accordingly, the whistler generates a drift displacement of electrons along the normal to the boundary with an amplitude
\begin{equation}
\xi_{\text{e}n1}=
\frac{\sigma_\text{ext\,IS}}{Zen_\text{i\,in}}
=-
\frac{\mathrm{i}\,|k_{n\,\text{ext\,IS}}|}{\varkappa_\text{in\,IS}}\,
\frac{\omega^2}{\omega_{B\text{i\,ext}}^2 }\,
\frac{n_\text{i\,ext}}{n_\text{i\,in}}\,
\xi_{\text{e}n0}
=-\sqrt{\sqrt{2}-1}\,\frac{\mathrm{i}\,|\omega|}{\omega_{B\text{i\,ext}}}\,
\frac{n_\text{i\,ext}}{n_\text{i\,in}}\,
\xi_{\text{e}n0}\ll \xi_{\text{e}n0}
\label{ElectronDisplacementInWhistler}
\end{equation}
(see formulas (\ref{SpatialDecrementOfIonSoundWaveInDenseHemiSpace}), (\ref{SquaredFrequencyOfIonSoundInDenseHemiSpace}), (\ref{SquaredNormalWaveNumberForOutgoingIonSound}), and (\ref{SurfaceChargeFromOutgoingIonSound})). We refine the magnetic pressure disturbance (\ref{MagneticPressureDisturbance}), taking into account
the additional electron displacement (\ref{ElectronDisplacementInWhistler}) in the full oscillation amplitude $\xi_{\text{e}n}=\xi_{\text{e}n0}+\xi_{\text{e}n1}$ of the magnetic field frozen into the electrons. We also take into account that the whistler does not disturb the electron density, and therefore the pressure in the dense half-space, due to the high thermal velocity of electrons compared to the whistler phase velocity. Then the pressure balance in the form of equality (\ref{PressureBalanceAsAdditionalSpectralRelation}) is corrected by multiplying its left-hand side by the correction factor $1+\xi_{\text{e}n1}/\xi_{\text{e}n0}$:
\begin{equation}
2\,|k_{\parallel}|\,\varkappa_\text{in\,IS}\,\frac{c_\text{s}^2}{\omega^2}
\,\Bigl(1+\frac{\xi_{\text{e}n1}}{\xi_{\text{e}n0}}\Bigr)=1.
\label{FineTunedPressureBalanceAsAdditionalSpectralRelation}
\end{equation}

The equality (\ref{FineTunedPressureBalanceAsAdditionalSpectralRelation}) is a dispersion relation with attenuation of the wave structure, if we take into account the equalities (\ref{SpatialDecrementOfIonSoundWaveInDenseHemiSpace}) and (\ref{SquaredFrequencyOfIonSoundInDenseHemiSpace}). The wavenumber $k_{\parallel}$ is kept real, and the frequency $\omega$ and the parameter $\varkappa_\text{in\,IS}$ are taken to be complex quantities with a small imaginary part $\Delta\omega$ and $\Delta\varkappa_\text{in\,IS}$ compared to the real part of the same quantities and proportional to the ratio $\xi_{\text{e}n1}/\xi_{\text{e}n0}\ll1$. The frequency $\omega$ and the parameter $\varkappa_\text{in\,IS}$ are still related to each other by the dispersion relation for ionic sound (\ref{DispersionRelationForIonSound}), (\ref{ImaginaryNormalComponentOfIonSoundWaveVector}). Therefore, the imaginary parts of $\Delta\omega$ and $\Delta\varkappa_\text{in\,IS}$ satisfy the equality $\omega\,\Delta\omega=-c_\text{s}^2\varkappa_\text{in\,IS}\,\Delta\varkappa_\text{in\,IS}$, or in equivalent form $(\Delta\varkappa_\text{in\,IS})/\varkappa_\text{in\,IS}=-(2\,\Delta\omega)/\omega$ taking into account the equalities (\ref{SpatialDecrementOfIonSoundWaveInDenseHemiSpace}) and (\ref{SquaredFrequencyOfIonSoundInDenseHemiSpace}).

Keeping in the equality (\ref{FineTunedPressureBalanceAsAdditionalSpectralRelation}) the first-order quantities with respect to the ratio $\xi_{\text{e}n1}/\xi_{\text{e}n0}$, we find the imaginary part of frequency
\begin{equation}
\delta\omega=\frac{\xi_{\text{e}n1}}{4\xi_{\text{e}n0}}\,\frac{\omega}{4}
=-\frac{\sqrt{\sqrt{2}-1}}{4}\,\frac{\mathrm{i}\,\omega^2}{\omega_{B\text{i\,ext}}}\,
\frac{n_\text{i\,ext}}{n_\text{i\,in}}\ll\omega,
\label{DecrementOfSurfaceStructure}
\end{equation}
which is significantly lower in absolute value than the real part of~$\omega$. This circumstance justifies the high quality factor of the considered surface structure.\,\footnote{\ The same order of magnitude of the decrement (\ref{DecrementOfSurfaceStructure}) is obtained as the ratio of the energy flux in the outgoing ion sound (along the normal to the boundary) to the energy in the pressed ion-sound mode in the dense half-space per unit area of the interface of the media.}

\subsection*{CONCLUSIONS}

Our analytical study explains the pronounced surface structure of the electric field along the upper and lower vaults of the magnetic arch, as revealed by numerical simulations~\cite{Korzhimanov-2025} of the collision of two opposing supersonic flows in a magnetic tube, where the dynamic pressure of each flow is chosen to be on the order of the pressure of the unperturbed magnetic field. The plasma flows expel the original magnetic field, resulting in a system with a high plasma pressure compared to the pressure of the magnetic field remaining in the tube. The plasma rope is held from further expansion by the pressure of the expelled magnetic field.

The surface structure is realized in the wave range between the ion gyrofrequencies $\omega_{B\text{i\,in}}$ and $\omega_{B\text{i\,ext}}$ inside and outside the arch. In this range, in the dense inner and rarefied outer medium, waves exist where the ionic and electron polarizations nearly compensate each other: ionic sound in a medium with a plasma pressure higher than the magnetic pressure, and magnetic sound in a medium with low intrinsic pressure. In ionic sound, electrons frozen into the magnetic field deform the magnetic field lines due to their thermal pressure and drift behind the unmagnetized ions. In magnetic sound, ions and electrons drift together: both fractions are frozen into the magnetic field. In a compressed ion sound, the expansion of the ion fraction across the magnetic field, pushing the intermedia boundary outward, is accompanied by stronger compression in the longitudinal direction ($k_{\parallel}^2>\varkappa_{\text{in\,IS}}^2$) and an increase in plasma pressure from the electrons maintaining quasi-neutrality. This boundary movement compresses the strong magnetic field from the outside, compensating for the variable pressure of the dense plasma by its own pressure.

A quasi-static linear transformation of ionic sound into magnetic sound occurs in the transition layer between the media under conditions of a uniform electron drift velocity $cE_\tau/B$. This regime gives rise to a significant increase in the tangential component of the electric field upon transitioning out of the arch~--- by $\omega_{B\text{i\,ext}}/\omega_{B\text{i\,in}}\gg1$ times. As a result, the electric component $E_\tau$ outside exceeds not only the similar weak component in the ``internal'' ionic sound, but also the total electric field in the latter~--- by $\omega_{B\text{i\,ext}}/\omega\gg1$ times. It is precisely this increase in electric field strength that attracted our attention in the numerical simulation~\cite{Korzhimanov-2025}.

This linear transformation of ionic sound into magnetic sound changes the polarization of the wave process in space. At the resonant surface within the transition layer, where the local ion gyrofrequency coincides with the wave process frequency, the electric field is strictly circularly polarized in a plane transverse to the magnetic induction, with the same direction of rotation as an electron. This polarization eliminates cyclotron resonance with cold ions.

The low polarization of the medium in magnetic sound (due to the combined drift of plasma fractions) allows it to match the ion sound at the boundary not only in pressure, but also in the normal component of electric induction (polarization). This ``double'' matching leaves only an energetically weak outgoing ion sound into the external rarefied medium (generated by the non-zero longitudinal component of the electric field in the pressed ion-acoustic mode). The outgoing wave causes slow attenuation of the wave structure. The runaway ion sound is close to electrostatic ion cyclotron oscillations almost perpendicular to the magnetic field. The stronger normal-to-the-boundary component of the electric field $E_n$ in such oscillations significantly exceeds the field of the pressed ion-acoustic mode in a dense plasma and reaches the level of the tangential component $E_\tau$ in magnetic sound. The $E_n$ component of the outgoing ionic sound oscillates on a smaller spatial scale than the localization region of the magnetic sound monotonically decaying from the boundary. The identical order of magnitude of the electrical components $E_\tau$ and $E_n$ corresponds to the results of the numerical simulation discussed, as do the smaller-scale oscillations of the normal component $E_n$.

The phase velocity of the surface wave under consideration is only approximately 10~\% lower than the phase velocity of the bulk ion sound (see formula (\ref{SquaredFrequencyOfIonSoundInDenseHemiSpace})). This circumstance does not create a specific prerequisite for the preferential buildup or attenuation of the surface wave compared to the bulk sound due to the Cherenkov resonance with electrons or ions characterized by a two-humped distribution with respect to longitudinal velocity. At the same time, the characteristic width $1/\kappa_\text{in\,IS}$ of the localization of surface ion sound in the plasma is relatively narrow and amounts to approximately one-third of the wavelength $2\pi/|k_{\parallel}|$. This characteristic significantly expands the range of ion velocities capable of ``resonantly'' interacting with the surface wave during its flight inside it, if the ion thermal velocity approaches the speed of sound. Therefore, the mechanism of surface wave buildup during the collision of dense supersonic plasma flows in a magnetic tube remains open.

The study is supported by the Russian Science Foundation (project No. 23-12-00317 ``Interaction of supersonic plasma flows in magnetic arch'').

\printbibliography

\clearpage

\makebox[0.99\textwidth]{\includegraphics{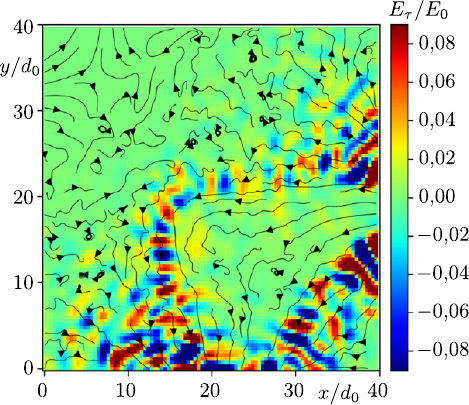}}
\noindent Fig.\ 1.\ Surface waves along the upper and lower vaults of the arch in the numerical simulation \cite{Korzhimanov-2025} in the case of injection of only one plasma flow. The color scheme depicts the tangential component of the strength $E_\tau$ at the time $t=60~\mu\text{s}$ from the start of injection of the flow with the density $n_\text{i\,foot}=10^{16}~\text{cm}^{-3}$ and the hydrodynamic velocity $v_\text{i}=10~\text{km}/\text{s}$ from the middle of the right boundary of the square calculation region. The field strength $E_\tau$ is normalized to the value $E_0=v_\text{i}B_0/c=25~\text{V}/\text{cm}$, where $B_0=0{.}25~\text{T}$ is the induction of external magnets at the injection point. The distances along the axes are normalized to the ionic inertial length $d_0=c/\omega_\text{pi}=1{.}2~\text{cm}$ in the incoming flow of singly ionized aluminum. Curves with arrows are magnetic field lines.

\clearpage

\makebox[0.99\textwidth]{\includegraphics{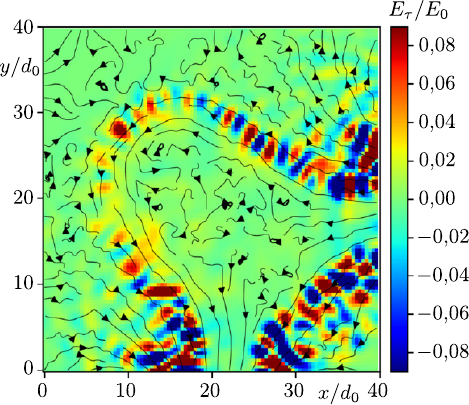}\includegraphics{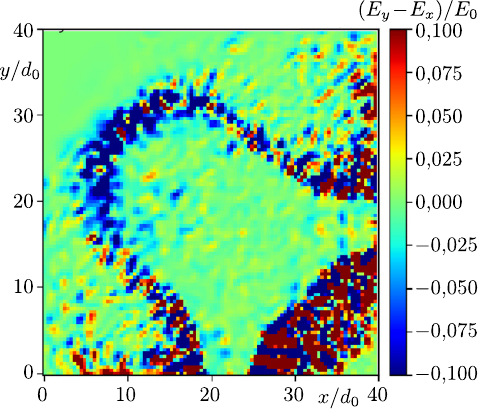}}
\noindent Fig.\ 2.\ Surface waves in the case of plasma flow injection from both bases of the arch. Parameters of each flow and external magnetic induction are the same as in Fig. 1. The right panel has been added with the ``normal'' component of the electric field strength along the normal $\mathbf{n}_\text{top}^0,\mathbf{E})/E_0$ at the top of the arch.

\makebox[0.99\textwidth]{\includegraphics{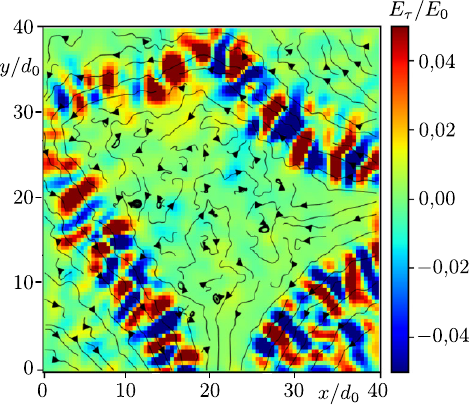}\includegraphics{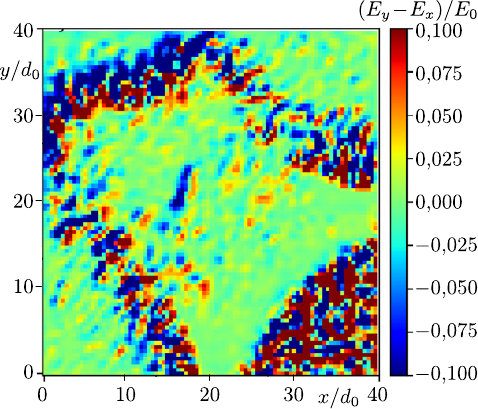}}
\noindent Fig.\ 3.\ Surface waves in the case of injection of plasma flows from both bases of the arch with a hydrodynamic velocity doubled $v_\text{i}=20~\text{km}/\text{s}$~\cite{Korzhimanov-2025}. Other parameters as in Fig.\ 2.

\end{document}
\typeout{get arXiv to do 4 passes: Label(s) may have changed. Rerun}